\documentclass[a4paper,11pt]{article}
\usepackage{pos}
\usepackage{subfig}
\usepackage{diagbox}

\newcommand{\Ddag}{\ensuremath{D^{\dag}} }
\newcommand{\cD}{\ensuremath{\mathcal D} }
\newcommand{\cDbar}{\ensuremath{\overline{\mathcal D}} }
\newcommand{\cF}{\ensuremath{\mathcal F} }
\newcommand{\cFbar}{\ensuremath{\overline{\mathcal F}} }
\newcommand{\Ibb}{\ensuremath{\mathbb I} }
\newcommand{\bn}{\ensuremath{\mathbf n} }
\newcommand{\cN}{\ensuremath{\mathcal N} }
\newcommand{\cO}{\ensuremath{\mathcal O} }
\newcommand{\pq}{\ensuremath{\mathrm{pq}} }
\newcommand{\cQ}{\ensuremath{\mathcal Q} }
\newcommand{\sB}{\ensuremath{s_{\text{Bos}}} }
\newcommand{\SB}{\ensuremath{S_{\text{Bos}}} }
\newcommand{\cU}{\ensuremath{\mathcal U} }
\newcommand{\cUbar}{\ensuremath{\overline{\mathcal U}} }
\newcommand{\be}{\ensuremath{\beta} }
\newcommand{\la}{\ensuremath{\lambda} }
\newcommand{\lalat}{\ensuremath{\la_{\text{lat}}} }
\newcommand{\vev}[1]{\ensuremath{\left\langle #1 \right\rangle} }
\newcommand{\pf}{\ensuremath{\text{pf}\,} }
\newcommand{\eq}[1]{Eq.~\ref{#1}}
\newcommand{\fig}[1]{Fig.~\ref{#1}}
\newcommand{\tab}[1]{Table~\ref{#1}}
\newcommand{\refcite}[1]{Ref.~\cite{#1}}
\newcommand{\secref}[1]{Section~\ref{#1}}

\title{Investigations of supersymmetric Yang--Mills theories}

\author*{Angel Sherletov}
\author{David Schaich}

\affiliation{Department of Mathematical Sciences, University of Liverpool, \\ Liverpool L69 7ZL, United Kingdom}

\emailAdd{A.Sherletov@liverpool.ac.uk}
\emailAdd{david.schaich@liverpool.ac.uk}

\FullConference{38th International Symposium on Lattice Field Theory \\ 26--30 July 2021 \\ Zoom/Gather @ Massachusetts Institute of Technology}

\abstract{
We present new results from ongoing lattice investigations of supersymmetric Yang--Mills (SYM) theories in three and four space-time dimensions. First considering the maximally supersymmetric 3d theory with $Q = 16$ supercharges, we check that the fermion pfaffian is approximately real and positive, validating phase-quenched RHMC calculations.  We then initiate lattice studies of running couplings and non-perturbative \be functions for $Q = 16$ SYM in both 3d and 4d, using a simple scheme based on Creutz ratios.  Finally, we consider 3d SYM with $Q = 8$ supercharges, developing new software as a first step towards supersymmetric QCD.
}

\begin{document}
\maketitle

\section{Introduction} 
In recent years there has been considerable interest and progress in the use of lattice field theory to non-pertubatively regularize and analyze supersymmetric quantum field theories---see \refcite{Schaich:2018mmv} for a recent review.
This is a challenging area of research due to the explicit breaking of the super-Poincar\'e algebra caused by the lattice discretization of space-time.
One of the most profitable approaches to address this challenge has been to employ `twisted' reformulations of supersymmetric Yang--Mills (SYM) theories, which preserve a closed supersymmetry sub-algebra at non-zero lattice spacing and thereby enable the recovery of the correct continuum limit with little to no fine-tuning.
However, this approach is limited to SYM theories with $Q \geq 2^d$ supersymmetries in $d$ dimensions---see \refcite{Catterall:2009it} for a thorough review.
These twisted SYM theories with $Q$ supercharges in $d$ dimensions also serve as a starting point for quiver constructions of lattice supersymmetric QCD with $Q/2$ supercharges in $d - 1$ dimensions~\cite{Catterall:2015tta}.

In this proceedings we discuss ongoing lattice investigations of three twisted SYM theories: 3d SYM with $Q = 8$ and $Q = 16$, as well as $Q = 16$ SYM in 4d.\footnote{The 4d $Q = 16$ theory is the famous $\cN = 4$ SYM.  Due to the nature of spinors in lower dimensions, the 3d theories with $Q = 8$ and $16$ are respectively $\cN = 4$ and $\cN = 8$ SYM.  For clarity we will label theories by the number of supersymmetries, $Q$, rather than by $\cN$.}
There are various motivations for considering these various theories.
The $Q=16$ theories are believed to be holographically dual to quantum gravity in $d + 1$ space-time dimensions, with the more modest computational costs of 3d $Q = 16$ SYM making it a promising target for non-trivial tests of holography~\cite{Catterall:2020nmn}.
Continuum 4d $Q=16$ SYM is the conformal field theory of the original AdS/CFT correspondence, while its 3d counterpart is believed to flow to a conformal IR fixed point, tying in to current interest in lattice studies of near-conformality for physics beyond the standard model~\cite{USQCD:2019hee}.
Finally, the 3d $Q = 8$ theory provides a starting point to analyze quiver super-QCD in two dimensions, as a prelude to 3d super-QCD based on 4d $Q=16$ SYM.

The following three sections address each of these topics in turn.
We begin by considering 3d $Q = 16$ SYM in the next section, improving some results from \refcite{Catterall:2020nmn} and analyzing the pfaffian.
In \secref{creutz} we discuss the running couplings for both $Q = 16$ theories, based on Creutz ratios.
Then, in \secref{3dQ8} we present newly developed software for 3d $Q = 8$ SYM, which is being added to the package presented by \refcite{Schaich:2014pda}.
We conclude in \secref{future} by briefly discussing the next steps for these projects.

\section{3d SYM with $Q = 16$: Dual black branes and pfaffian phase}\label{3dQ16} 
\begin{figure}[tbp]
  \includegraphics[width=0.47\linewidth]{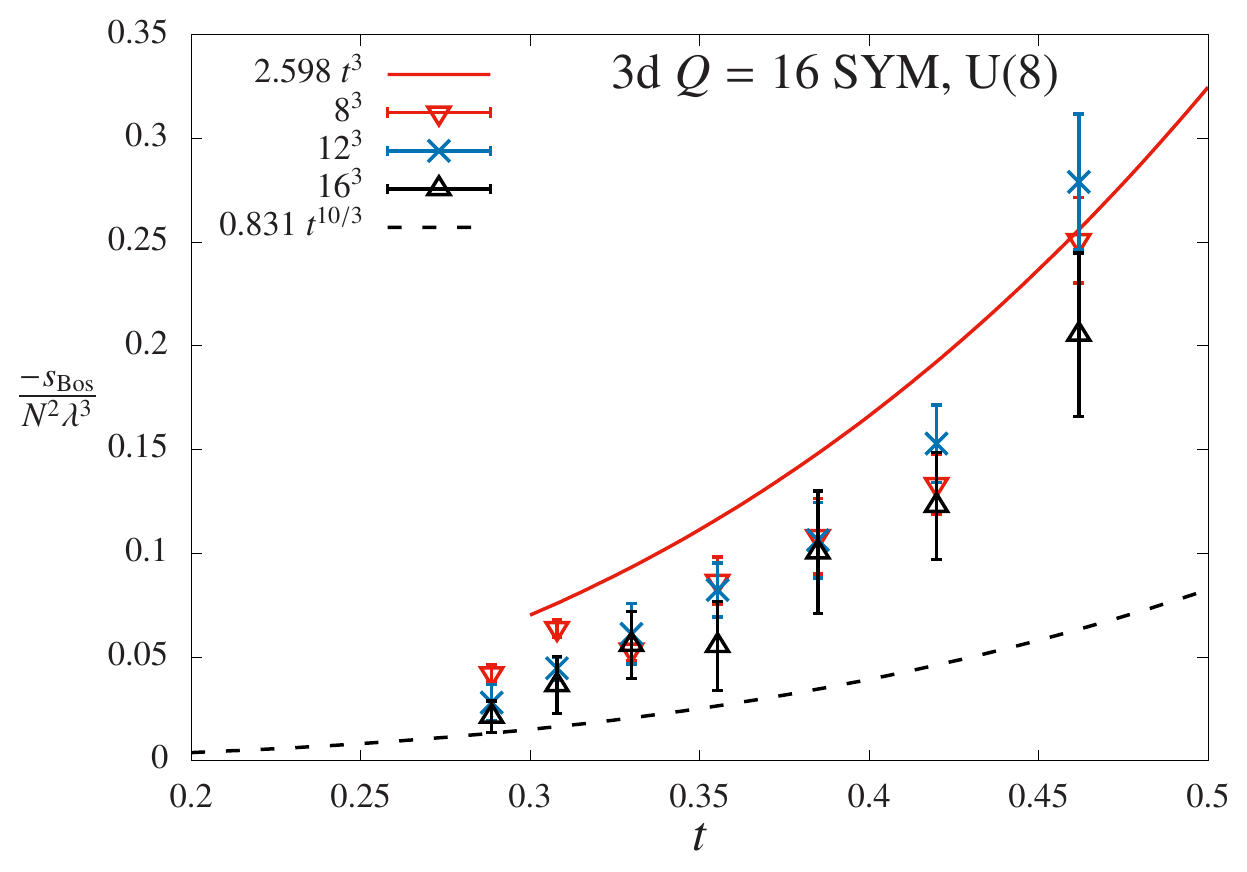}
  \hfill
  \includegraphics[width=0.47\linewidth]{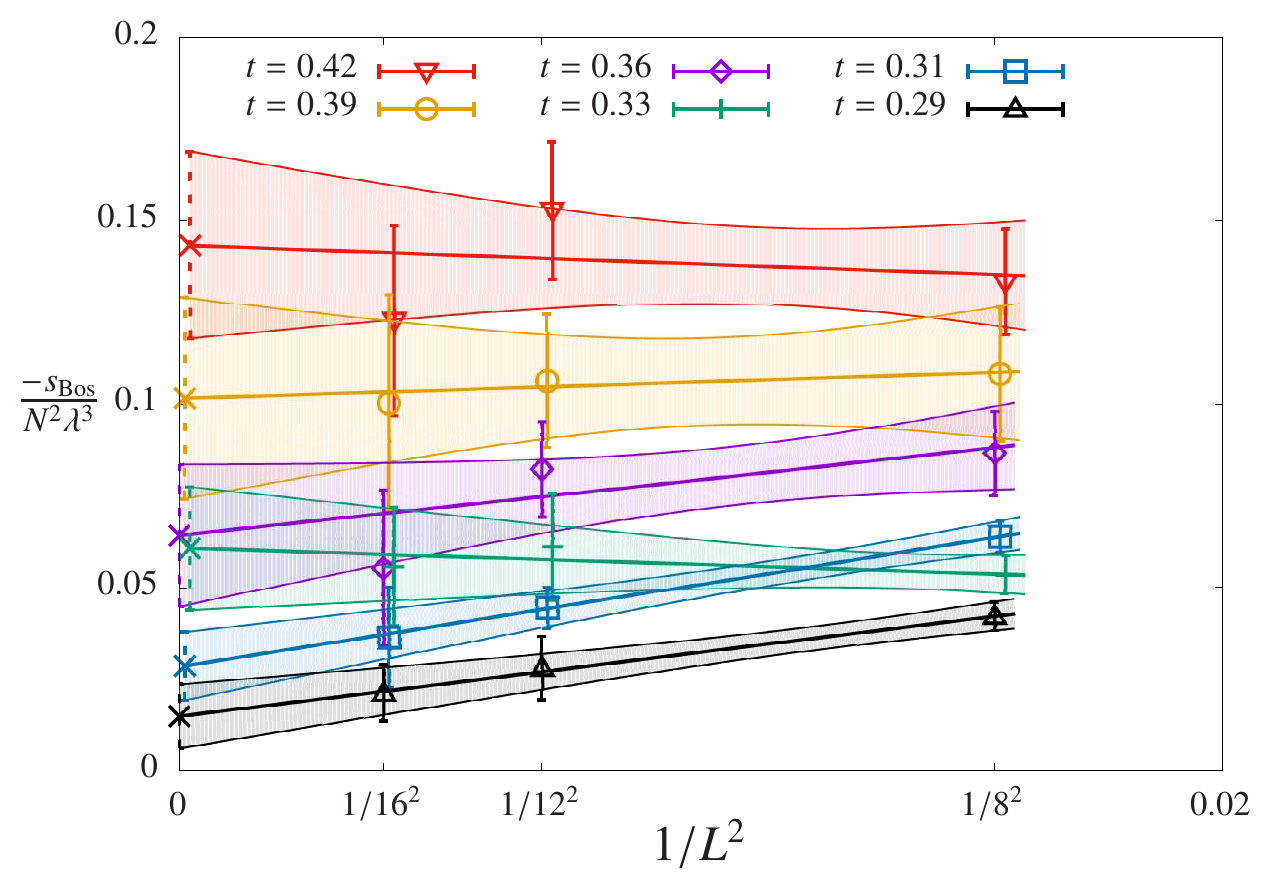}
  \caption{Bosonic action density for 3d $Q = 16$ SYM with gauge group U(8) and $L^3$ lattice volumes with $L = 8$, $12$, $16$, improving upon Figs.~1 and 4 in \refcite{Catterall:2020nmn} by adding new data.  \textbf{Left:} All results vs.\ the dimensionless temperature $t$, comparing with the high-temperature expectation $\propto$$t^3$ and low-temperature dual-supergravity prediction $\propto$$t^{10 / 3}$.  \textbf{Right:} Linear $L^2 \to \infty$ continuum extrapolations for the six lowest $t$.}
  \label{fig:3dQ16}
\end{figure}

The first numerical lattice studies of $Q = 16$ SYM in three dimensions were recently carried out~\cite{Catterall:2020nmn}, focusing on the behavior of dual black D$2$-branes at finite temperature.
This work employs a straightforward dimensional reduction of the 4d lattice theory, using the 4d code presented in \refcite{Schaich:2014pda} to consider $N_s^2 \times 1 \times N_{\tau}$ lattice volumes with the aspect ratio $N_s / N_{\tau} = 1$ chosen to correspond to the homogeneous D$2$-phase of the dual gravitational theory.
The 't~Hooft coupling \la of this continuum 3d theory is dimensionful, and can be combined with the dimensionful temperature $T = 1 / (aN_{\tau})$ to define a dimensionless temperature in terms of a dimensionless lattice 't~Hooft coupling:
\begin{align}
  t & = \frac{\sqrt{3} T}{4\la} = \frac{1}{\lalat N_{\tau}} &
  \lalat & = \frac{(d+1)^{\frac{5-d}{8-2d}}}{\sqrt{d}} a^{4-d} \la = \frac{4}{\sqrt{3}} a\la,
\end{align}
where the numerical factor arises from discretizing the theory on the $A_d^*$ lattice in $d$ dimensions~\cite{Catterall:2017lub}.

One observable of interest is the bosonic action density $\sB$, which corresponds to the free energy density of the dual supergravity at low temperatures in the large-$N$ limit of the SU($N$) gauge group.
A notable achievement of \refcite{Catterall:2020nmn} was carrying out the first $N_{\tau} \to \infty$ continuum extrapolations of $\sB$, which send $\lalat \to 0$ in order to keep the dimensionless temperature $t$ fixed.
In the time since \refcite{Catterall:2020nmn} was published, we have accumulated more data for several of the ensembles it analyzed, leading to the improved results presented in \fig{fig:3dQ16}.
These results for gauge group U(8) show qualitative agreement emerging at low temperatures between the lattice calculations and the large-$N$ dual-gravity black brane prediction.

\begin{figure}[tbp]
  \centering
  \includegraphics[width=0.47\linewidth]{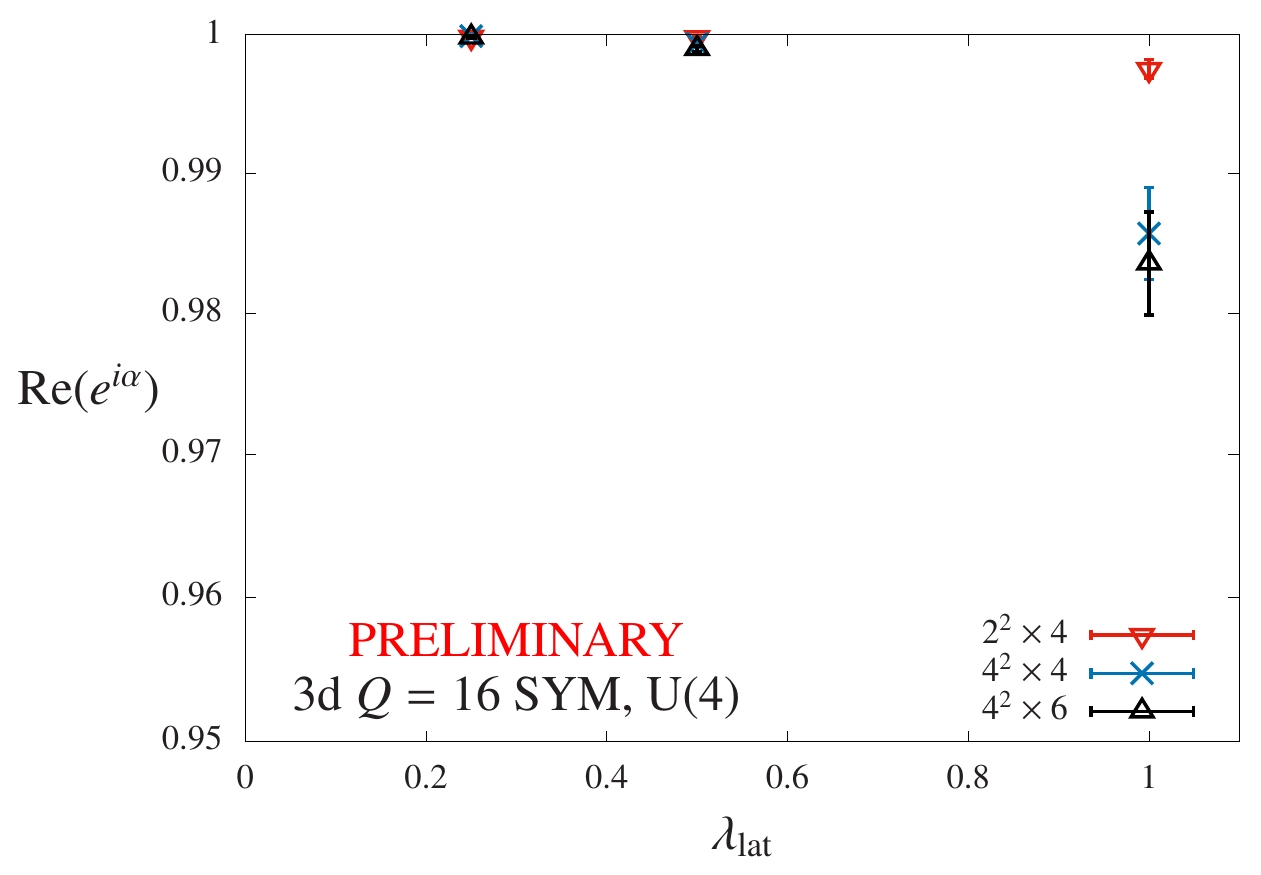}
  \caption{Pfaffian phase fluctuations vs.\ \lalat for 3d $Q = 16$ SYM with gauge group U(4) and volumes up to $4^2 \times 6$, establishing that there is no sign problem in the $N_{\tau} \to \infty$ continuum limit with $\lalat \to 0$.}
  \label{fig:pfaffian}
\end{figure}

In addition, we have further validated the results from \refcite{Catterall:2020nmn} by investigating the complex pfaffian that results from integrating over the fermion fields of the theory, $\int \left[d\Psi\right] e^{-\Psi^T \cD \Psi} \propto \pf \cD = |\pf \cD| e^{i\phi}$.
This complex weight in the path integral obstructs importance sampling approaches such as the rational hybrid Monte Carlo (RHMC) algorithm used in this work, which we address by `quenching' the phase $e^{i\phi} \to 1$.
This formally requires evaluating the phase-quenched (pq) expectation value $\vev{e^{i\phi}}_{\pq}$ in order to perform phase reweighting:
\begin{align}
  \vev{\cO} & = \frac{\vev{\cO e^{i\phi}}_{\pq}}{\vev{e^{i\phi}}_{\pq}} &
  \text{where} & &
  \qquad \vev{\cO}_{\pq} & = \frac{\int[d\cU][d\cUbar] \, \cO \, e^{-\SB} |\pf \cD|}{\int[d\cU][d\cUbar] \, e^{-\SB} |\pf \cD|}.
\end{align}
However, the calculation of the pfaffian phase is far more computationally demanding than RHMC configuration generation,\footnote{For example, each $4^2\times 6$ pfaffian measurement going into \fig{fig:pfaffian} took nearly $500$ core-hours, compared to less than a single core-minute for each RHMC trajectory.} making it impractical to compute $\vev{e^{i\phi}}_{\pq}$ for the U(8) gauge group and large volumes up to $16^3$ considered in \refcite{Catterall:2020nmn}.

Here, instead, we consider smaller volumes up to $4^2 \times 6$ with a smaller U(4) gauge group, obtaining the results shown in \fig{fig:pfaffian}.
This figure plots the real part of $\vev{e^{i\phi}}_{\pq}$, which deviates only slightly from unity due to small fluctuations around the positive real axis.
As seen for 4d $Q = 16$ SYM~\cite{Schaich:2018mmv}, these phase fluctuations grow as the volume increases, but shrink as \lalat decreases.\footnote{No dependence was seen on the rank of the gauge group~\cite{Schaich:2018mmv}.}
The question is which of these effects will win in the $N_{\tau} \to \infty$ continuum limit with $\lalat \to 0$.
From \fig{fig:pfaffian} we can see that the larger fluctuations resulting from quadrupling the volume from $2^2 \times 4$ to $4^4\times 4$ for are easily compensated by reducing $\lalat = 1 \to 0.25$.
This provides reassurance that $\vev{e^{i\phi}}_{\pq} \approx 1$ for the larger volumes considered in \refcite{Catterall:2020nmn}, even the smallest of which has $\lalat \leq 1$.

\section{Running couplings from Creutz ratios in 3d and 4d}\label{creutz} 
We turn now to considering near-conformal dynamics, which connects SYM to broader areas of research into fundamental aspects of quantum field theory and potential new physics beyond the standard model, reviewed in \refcite{USQCD:2019hee}.
By studying $Q = 16$ SYM, we have the advantage of knowing that the continuum theory is either exactly conformal for all 't~Hooft couplings $\la = g^2 N$ (in 4d) or is believed to flow to a conformal IR fixed point (in 3d).
In the spirit of \refcite{Bergner:2021ffz}, we can therefore use lattice studies of running couplings and anomalous dimensions to explore the effects of explicit conformal symmetry breaking from the non-zero lattice spacing and finite lattice volume.

Here we focus on running couplings defined through a simple lattice scheme introduced by \refcite{Bilgici:2009kh} and based on the Creutz ratio~\cite{Creutz:1980wj}
\begin{equation*}
  \chi(I,J) = -\log\frac{W(I,J)W(I-1,J-1)}{W(I-1,J)W(I,J-1)},
\end{equation*}
where $W(I,J)$ is the trace of the rectangular $I\times J$ Wilson loop averaged over orientations and the lattice volume.
The twisted formulation of SYM introduces complexified gauge links $\cU_a$, and we construct the Wilson loops out of the unitary parts $U_a$ of these complexified links, extracted through the polar decomposition $\cU_a(\bn) = e^{\phi_a(\bn)} U_a(\bn)$.
Considering the relative scale $r=\hat{R}/{N}=\hat{R}a/L$ within lattice length $L = aN$, and denoting by $\tilde{\chi}(N,L|\hat{R})$ the Creutz ratio $\chi(\hat{R},\hat{R})$ measured on an $N^4$ lattice with lattice spacing $a = L / N$, we can define a running coupling
\begin{equation*}
  k(r) g^2(r,N,L) \equiv \hat{R}^2\tilde{\chi}(N,L|\hat{R})
\end{equation*}
up to an $r$-dependent factor $k$.\footnote{On small $N^4$ lattice volumes, $k$ may also depend on $N$~\cite{Bilgici:2009kh, Giedt:2011kz}, a possible systematic effect that we don't explore here.}
While this approach has been surpassed by gradient-flow methods for QCD-like theories~\cite{Ramos:2015dla}, its simplicity provides a useful starting point for novel systems such as $Q = 16$ SYM.

The scale $r$ is part of the renormalization scheme, so in order to explore the scale dependence of $kg^2$ we need to consider several $N^4$ lattice volumes with fixed $r \leq 0.5$, which also fixes the factor $k(r)$.
In general $\hat R = rN$ will not be an integer, which requires interpolating between results for $I\times I$ and $(I + 1) \times (I + 1)$ Creutz ratios with $I < \hat R < I + 1$.
This interpolation introduces potentially significant systematic uncertainties, which we remain in the process of estimating and will omit from this proceedings.
We also need to keep the lattice spacing $a = L / N$ fixed.
For 4d $Q = 16$ SYM we assume that this can be done by fixing the lattice 't~Hooft coupling \lalat defined at the scale of the lattice spacing.
For the 3d case we similarly fix the dimensionless temperature $t$ discussed in \secref{3dQ16}.

\begin{table}[btp]
  \centering
  \begin{tabular}{|c|c|c|c|c|c|}
    \hline
    \diagbox[width=1.2cm, height=0.7cm]{$r$}{$N$} & 8      & 10     & 12     & 14     & 16     \\\hline
    0.3                                           & 0.0891 & 0.0591 & 0.0427 & 0.0491 & 0.0442 \\\hline
    0.4                                           & 0.0551 & 0.0415 & 0.0233 & 0.0362 & 0.0315 \\\hline
  \end{tabular}
  \caption{$k g^2$ for 4d $Q = 16$ SYM with gauge group U(2) and $\lalat=0.5$, considering two values of $r$ and five $N^4$ lattice volumes.}
  \label{tab:4d}
\end{table}

\begin{table}[btp]
  \centering
  \subfloat[$t=0.31$]{
    \begin{tabular}{|c|c|c|c|}
      \hline
      \diagbox[width=1.2cm, height=0.7cm]{$r$}{$N$} & 8     & 12    & 16    \\\hline
      0.3                                           & 0.385 & 0.240 & 0.170 \\\hline
      0.4                                           & 0.270 & 0.180 & 0.131 \\\hline
    \end{tabular}
  }
  \qquad\qquad\qquad
  \subfloat[$t=0.42$]{
    \begin{tabular}{|c|c|c|c|}
      \hline
      \diagbox[width=1.2cm, height=0.7cm]{$r$}{$N$} & 8     & 12    & 16    \\\hline
      0.3                                           & 0.320 & 0.204 & 0.165 \\\hline
      0.4                                           & 0.307 & 0.205 & 0.182 \\\hline
    \end{tabular}
  }
  \caption{$kg^2$ for 3d $Q = 16$ SYM with gauge group U(8) and two dimensionless temperatures, in each case considering two values of $r$ and three $N^3$ lattice volumes.} 
  \label{tab:3d}
\end{table}

In \tab{tab:4d} we collect some results for $kg^2$ from several 4d $Q = 16$ SYM lattice ensembles recently analyzed by \refcite{Bergner:2021ffz}.
These ensembles have gauge group U(2), $\lalat = 0.5$, and lattice volumes $8^4$ through $16^4$.
Similarly, in \tab{tab:3d} we collect some results from several 3d $Q = 16$ SYM ensembles for two of the dimensionless temperatures ($t = 0.31$ and $0.42$) shown in \fig{fig:3dQ16}.
In each case we consider $r = 0.3$ and $0.4$ chosen to reduce both small-$r$ discretization artifacts and large-$r$ finite-volume artifacts.
Statistical uncertainties on these results are negligible, while we are still working to estimate systematic uncertainties including those from interpolating between the $r \propto 1 / N$ directly accessible on $N^4$ lattices.
We therefore omit uncertainties entirely in these tables.

Comparing Tables~\ref{tab:4d} and \ref{tab:3d}, we can see an interesting qualitative contrast between the 4d and 3d maximally supersymmetric theories.
While these numerical values for $kg^2$ are not directly meaningful due to the unknown factor $k$, the significantly smaller results for 4d compared to 3d with similar $N$ and \lalat are striking.
(Any effects of the different gauge group should cancel out in the Creutz ratio.)
In particular, depending on the as-yet unknown systematic errors, the 4d results for each $r$ may be constant within these uncertainties for $L \gtrsim 10$, which would be consistent with the conformality of the continuum theory.
The 3d results show more significant decreases as $L$ increases, which could be a first sign of `backward running' towards an IR fixed point.

\begin{table}[btp]
  \centering
  \begin{tabular}{|c|c|c|c|c|c|}
    \hline
    \diagbox[width=1.2cm, height=0.7cm]{$r$}{$N$} & 8      & 10     & 12     & 14     & 16     \\\hline
    0.3                                           & 0.0922 & 0.0645 & 0.0625 & 0.0537 &  0.0513 \\\hline
    0.4                                           & 0.0588 & 0.0490 & 0.0462 & 0.0431 &  0.0403 \\\hline
  \end{tabular}
  \caption{$k g^2$ for 4d $Q = 16$ SYM with gauge group U(2), $\lalat=0.5$ and fermion mass $m_f=0.25$, considering two values of $r$ and five $N^4$ lattice volumes.}
  \label{tab:4d_fmass}
\end{table}

In order to clarify the interpretation of these results, we have begun generating additional lattice ensembles with non-zero fermion masses $m_f$---explicitly breaking the lattice supersymmetry by shifting the fermion operator $\Ddag D \to \Ddag D + m_f^2 \Ibb$.
As the fermion mass increases, the system approaches a gauge--scalar theory with a rapidly running coupling, providing a benchmark for comparison with Tables~\ref{tab:4d} and \ref{tab:3d}.
This strategy was previously used by \refcite{Giedt:2011kz}.
\tab{tab:4d_fmass} presents some initial 4d results with $m_f = 0.25$, which feature systematically larger $kg^2$ compared to \tab{tab:4d} but don't qualitatively change the potentially conformal behavior.
Separately, we are also working to adapt gradient-flow running coupling methods to twisted lattice SYM, which we hope will improve upon this initial analysis from Creutz ratios.

\section{3d SYM with $Q = 8$: Towards super-QCD}\label{3dQ8} 
Finally we report on our work developing new parallel software for 3d $Q = 8$ SYM, which we are carrying out as a first step towards reproducing and surpassing the only prior numerical lattice study of 2d quiver super-QCD in \refcite{Catterall:2015tta}.\footnote{Earlier work developing lattice quiver super-QCD formulations includes Refs.~\cite{Matsuura:2008cfa, Sugino:2008yp, Kadoh:2009yf, Joseph:2013jya, Joseph:2013bra, Joseph:2014bwa}, while \refcite{Giedt:2017fck} presents an alternative twisted formulation of lattice 3d $Q = 8$ SYM.}
This system has been implemented within the publicly available package presented in \refcite{Schaich:2014pda}.\footnote{{\tt\href{https://github.com/daschaich/susy}{github.com/daschaich/susy}}}
Following Refs.~\cite{Catterall:2011cea, Catterall:2015tta}, we begin with the lattice action
\begin{multline}
  \label{eq:S}
  S = \frac{N}{4\lalat} \sum_{\bn} \mathrm{Tr} \Big[-\cFbar_{ab}(\bn)\cF_{ab}(\bn) + \frac{1}{2}\left(\cDbar^{(-)}_a \cU_a(\bn)\right)^2 - \chi_{ab}(\bn)\cD^{(+)}_{[a}\psi_{b]}(\bn) \\
  - \eta(\bn)\cD^{(-)}_a\psi_a(\bn) - \theta_{abc}(\bn)\cD^{(+)}_{[a}\chi_{bc]}(\bn)\Big],
\end{multline}
where $(\eta, \psi_a, \chi_{ab}, \theta_{abc})$ are the $1 + 3 + 3+ 1$ antisymmetric-tensor fermion field components.
In contrast to the $A_3^*$ (body-centered cubic) lattice required for the maximally supersymmetric theory, we discretize $Q = 8$ SYM on a simple cubic lattice, summing over all lattice sites $\bn$.
Repeated indices are summed over $1, \cdots, 3$.

Just as for 4d $Q = 16$ SYM~\cite{Catterall:2015ira}, we add two deformations to \eq{eq:S} in order to stabilize numerical calculations.
The first of these is a simple scalar potential that lifts the SU($N$) flat directions.
In our software we provide two different options for this scalar potential: 
\begin{align}
  \label{eq:pot}
  \frac{N}{4\lalat} \mu^2 \sum_{\bn, a} \left(\frac{1}{N}\mathrm{Tr}\left[\cU_a(\bn) \cUbar_a(\bn)\right]-1\right)^2 \qquad & & &
  \frac{N}{4\lalat} \mu^2 \sum_{\bn, a} \mathrm{Tr}\left[\bigg(\cU_a(\bn) \cUbar_a(\bn) - \Ibb_N\bigg)^2\right].
\end{align}
The first option constrains the trace of each $\cU_a(\bn) \cUbar_a(\bn)$ (no sum over $a$) while the second constrains each eigenvalue of $\cU_a(\bn) \cUbar_a(\bn)$ individually.
In practice both choices produce similar behavior, and both are currently in use (e.g., the first by Refs.~\cite{Catterall:2015tta, Bergner:2021ffz} and the second by Refs.~\cite{Catterall:2017lub, Catterall:2020nmn}).
In particular, both options fail to affect a U(1) phase that cancels out of the product $\cU_a(\bn) \cUbar_a(\bn)$.
A second deformation is required to lift flat directions in this U(1) sector, and following \refcite{Catterall:2015ira} we implement this supersymmetrically by modifying the moduli equations to depend on the determinant of the plaquette.

\begin{figure}[tbp]
  \centering
  \includegraphics[width=0.47\linewidth]{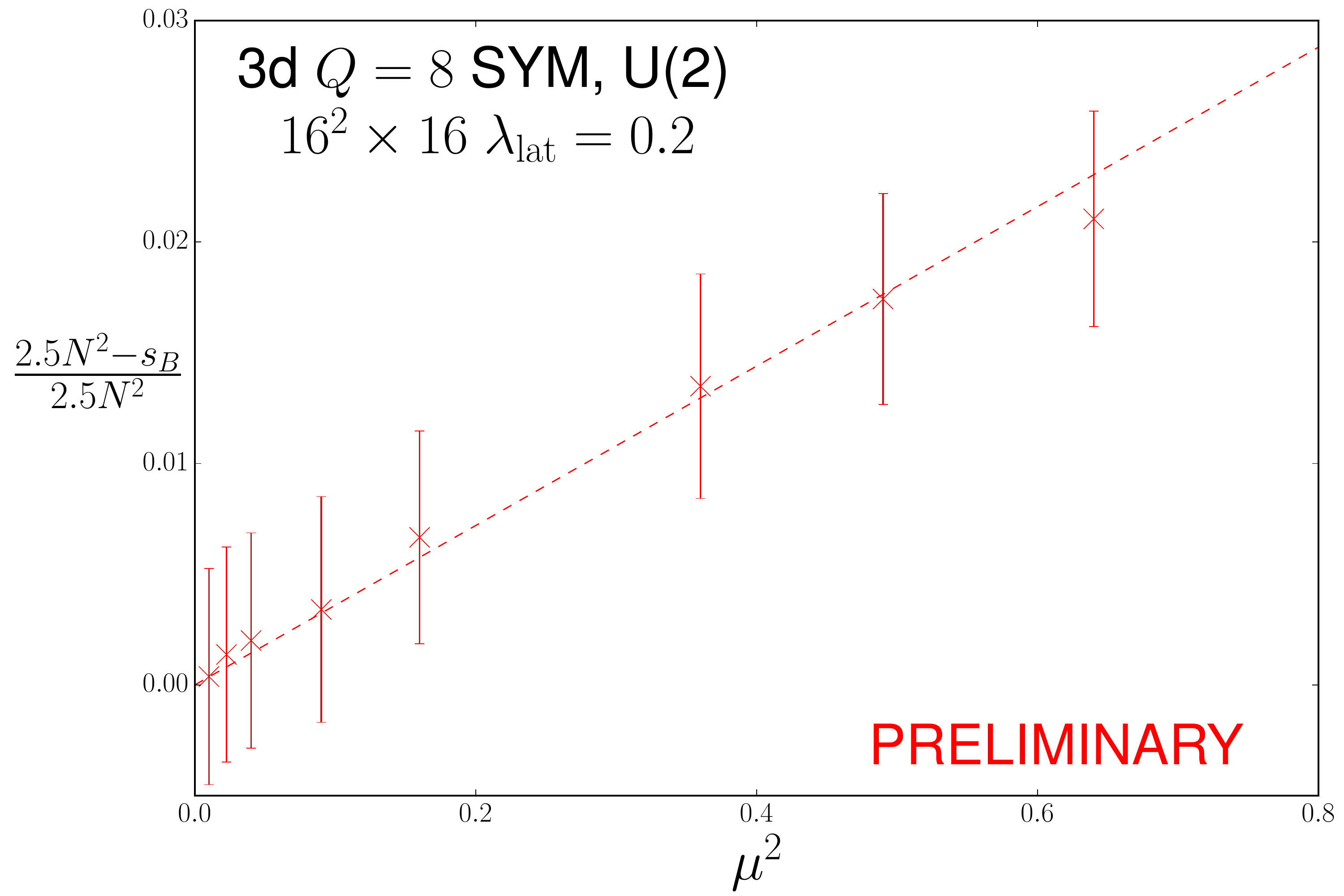}
  \caption{Violations of a \cQ supersymmetry Ward identity for 3d $Q = 8$ SYM, considering gauge group U(2) with lattice volume $16^3$ and $\lalat=0.2$.}
  \label{fig:3dQ8}
\end{figure}

Although the plaquette determinant deformation can be implemented supersymmetrically, both options for the scalar potential in \eq{eq:pot} softly break the single twisted supersymmetry \cQ that \eq{eq:S} preserves at non-zero lattice spacing.
In \fig{fig:3dQ8}, as a test of our implementation of 3d $Q = 8$ SYM, we plot violations of a \cQ supersymmetry Ward identity that fixes the exact value of the bosonic action density, $\sB = 5N^2 / 2$.
Here we use the first (double-trace) option in \eq{eq:pot}.
We can see that these violations vanish proportionally to the tunable parameter $\mu^2$ in the scalar potential, confirming that the \cQ supersymmetry is broken only softly and recovered in the $\mu^2 \to 0$ limit.

\section{Outlook and next steps}\label{future} 
In this proceeding we have presented new results from ongoing lattice investigations of three twisted SYM theories.
First, for 3d $Q = 16$ SYM we improved results from \refcite{Catterall:2020nmn} for the bosonic action density that is holographically dual to the free energy density of black D$2$-branes in supergravity, and also explicitly checked that pfaffian phase fluctuations for this theory appear to remain well under control in the continuum limit.
Then we discussed initial explorations of running couplings for $Q = 16$ SYM in both three and four dimensions, employing a simple lattice scheme based on Creutz ratios.
Although this work remains preliminary, we observed an interesting contrast between the 3d and 4d results, which could be consistent with the conformality of 4d $Q = 16$ SYM.
Finally we presented initial tests of newly developed parallel software for 3d $Q = 8$ SYM, confirming that violations of a supersymmetric Ward identity are proportional to the supersymmetry-breaking scalar potential required to stabilize numerical calculations.

There are clear next steps for each of these three projects, most of which are currently underway.
First, for 3d $Q = 16$ SYM we have begun large-scale computations with multiple $N_s / N_{\tau}$ aspect ratios, which will allow us to study non-perturbative phase transitions predicted by holography at low temperatures.
For the running coupling study, we are currently generating new lattice ensembles with non-zero fermion masses in an attempt to clarify our current results.
However, it may prove necessary to switch to more robust gradient-flow running coupling methods, which we are also working on.
Finally, the new 3d $Q = 8$ SYM software presented in \secref{3dQ8} is being developed as a first step towards numerical lattice studies of quiver super-QCD, first targeting the 2d theory considered by \refcite{Catterall:2015tta} and then building on that experience to begin investigating super-QCD in three dimensions.

\vspace{20 pt} 
\noindent \textsc{Acknowledgments:}~We thank Raghav Jha, Toby Wiseman, Anosh Joseph, Georg Bergner, Simon Catterall and Joel Giedt for helpful conversations and continuing collaboration on lattice supersymmetry.
Numerical calculations were carried out at the University of Liverpool.
DS was supported by UK Research and Innovation Future Leader Fellowship {MR/S015418/1} and STFC grant {ST/T000988/1}.

\bibliographystyle{JHEP}
\bibliography{lattice21}
\end{document}